\documentclass[doublecol]{epl2}
\usepackage{amssymb}
\usepackage{graphicx}

\title{Planck-scale effects on Bose-Einstein condensates}

\author{F. Briscese \inst{1,2} \thanks{E-mail: \email{briscese.phys@gmail.com}} \and
M. Grether \inst{3} \thanks{E-mail:
\email{mdgg@hp.fciencias.unam.mx}} \and M. de Llano \inst{4}
\thanks{E-mail: \email{dellano@unam.mx}}}

\shortauthor{F. Briscese \etal}

\institute{\inst{1} Escuela de F\'{\i}sica, Universidad Industrial
de Santander -
Ciudad Universitaria, Bucaramanga 680002, COLOMBIA\\

\inst{2} Istituto Nazionale di Alta Matematica Francesco Severi,
Gruppo Nazionale di Fisica Matematica - Piazzale Aldo Moro 5,
c.a.p. 00185, Rome, ITALY \\

\inst{3}Facultad de Ciencias, Universidad Nacional Aut\'{o}noma de
M\'{e}xico - 04510 M\'{e}xico, DF,MEXICO. \\

\inst{4} Instituto de Investigaciones en Materiales, Universidad
Nacional Aut\'{o}noma de M\'{e}xico - 04510 M\'{e}xico, DF,\\
MEXICO. }


\abstract{The effects of a Planck-scale deformation of the
Minkowski energy-momentum dispersion relation on the phenomenology
of non-trapped Bose-Einstein condensates (BECs) are examined. Such
a deformation is shown to cause a shift in the condensation
temperature $T_{c}$ of the BEC and, for a specific functional form
of deformation, this shift can be as large as the current measured
precision on $T_{c}$. For a $_{37}^{85}Rb$ cold-atom BEC with a
particle density $n\simeq 10^{12}cm^{-3}$ we find a fractional
shift of order $10^{-4}$, but this can be much larger for even
more dilute BECs. We discuss the possibility of planning specific
experiments with BECs that might provide phenomenological
constraints on Planck-scale physics. These corrections to $T_{c}$
are found to be extremely small for ultrarelativistic BECs
implying that, in some cases, Planck-scale effects may be more
important in low- rather than high-energy processes.  }

\begin{document}

\maketitle

Lively interest has recently emerged to experimentally probe
quantum-gravitational Planck-scale effects
\cite{amelinophenomenology} due to deformations of the standard
Minkowski free-particle energy-momentum dispersion relation of
special relativity. Such deformations are a general feature of
quantum-gravity theories, as e.g., loop quantum gravity
\cite{LQG,a3,LQG3,LQG4,LQG5,LQG6,LQG7} or noncommutative
geometries \cite{NCG,NCG4,NCG2,NCG3}. Further, it is also of great
relevance in the general context of Lorentz-symmetry breaking
\cite{LSB}. Quite generally, a deformed dispersion relation can be
written as%
\begin{equation}
\begin{array}{ll}
E(p)\equiv E_{0}(p)+\delta E(p,m,M_{P}) &
\end{array}
\label{deformedenergydensitydefinition0}
\end{equation}%
where $E_{0}(p)\equiv \sqrt{p^{2}c^{2}+m^{2}c^{4}}$ is the familiar
Minkowski dispersion relation, with $p$ the particle momentum, $m$ its rest
mass and $c$ the speed of light. Here $\delta E(p,m,M_{P})$ represents the
deviation from $E_{0}(p)$ which in addition is a function of the Planck mass
$M_{P}$ while its explicit form depends on the details of the
quantum-gravity model used. The relation (\ref%
{deformedenergydensitydefinition0}) is assumed to be universal, i.e., is the
same for all the elementary particles, including composite particles such as
nucleons or atoms when internal degrees of freedom are negligible.

The dependence on the Planck mass $M_{P}$\ is introduced since one
expects that departures from Lorentz symmetry become important at
Planck scales, but one should require that this symmetry be
restored when $p\ll cM_{P}$. Moreover one would like to preserve
the interpretation of $m$ as the particle rest mass. In general,
therefore, one should impose the two conditions

\begin{equation}
\begin{array}{ll}
\delta E(p=0,m,M_{P})=0\\
\\
\delta E(p,m,M_{P})_{\overrightarrow{M_{P}\longrightarrow \infty }}0. \\
\end{array}
\end{equation}
Initially, attempts to constrain the functional form of $\delta
E(p,m,M_{P})$ were in an astrophysical context where particles are
in the ultrarelativistic (UR) regime, $p\gg mc$
\cite{astrophysicalregime}. In this limit the dispersion relation
deformation can be parameterized quite generally, regardless of
the explicit model under study. Due to the extremely large value
of $M_{P}\simeq 10^{19}GeV/c^{2}$ a series expansion of $\delta
E(p,m,M_{P})$ in inverse powers of $M_{P}$ might prove useful
along with the leading term in $1/M_{P}$ as first approximation.
Accordingly, $\delta E(p,m,M_{P})$ would be written
\cite{amelino,am1}
as%
\begin{equation}
\delta E(p,m,M_{P})\simeq \frac{1}{2M_{P}}\left( \eta
_{1}p^{2}+\eta _{2}mcp+\eta _{3}m^{2}c^{2}\right)
\label{deltaEURparametrization}
\end{equation}%
with the three real parameters: $\eta _{1}$ associated with the
leading term, $\eta _{2}$ to the next leading term, and $\eta_{3}$
to the next-to-next leading term. As pointed out in many studies
\cite{astrophysicalregime,a1,a2,a4,a5}, astrophysical data could
be sensitive to a leading-order deformation with $|\eta_{1}|$
$\lesssim \mathbf{1}$. A preliminary analysis of the Fermi Space
Telescope data \cite{fermispacetelescope,f1,f2,f3,f4} is currently
underway to constrain $\eta _{1} $. Very recently
\cite{amelino,am1}, constraining the functional form of
(\ref{deformedenergydensitydefinition0}) in the nonrelativistic
(NR) regime $p\ll mc$ based on ultra-precise
cold-atom-recoil-frequency experiments was proposed. In the NR
limit, a more appropriate parametrization of $\delta E(p,m,M_{P})$
is \cite{amelino,am1}
\begin{equation}
\delta E(p,m,M_{P})\simeq \frac{1}{2M_{P}}\left( \xi _{1}mcp+\xi
_{2}p^{2}+\xi _{3}\frac{p^{3}}{mc}\right)  \label{deltaENRparametrization}
\end{equation}%
with parameter bounds obtained as $-6.0<\xi _{1}<2.4$ for the leading-order
deformation parameter and $-3.8\times 10^{9}<\xi _{2}<1.5\times 10^{9}$ for
the next-to-leading order, both within a $95\%$ confidence level \cite%
{amelino,am1}.

The low- and high-energy bounds just mentioned are perfectly
complementary. To illustrate this fact, consider the  following
deformation $\delta E(p,m,M_{P})=-\eta \left[
m^{3}c^{4}/\sqrt{m^{2}c^{4}+p^{2}c^{2}} -m^{2}c^{2}\right] /M_{P}$
\cite{amelino}. In the low-energy NR limit this dispersion
relation corresponds to (\ref{deltaENRparametrization}) with $\xi
_{2}=-\eta $, $\xi_{1}=\xi_{3}=0$, namely the next-to-leading
order. In the high-energy UR limit, on the other hand, this
corresponds to (\ref{deltaEURparametrization}) with $\eta
_{3}=2\eta $, $ \eta _{1}=\eta _{2}=0$ or to the next-to-next
leading order. This example also exhibits how the the deformation
$\delta E$ can be more important in the NR than in the UR limit.

It might be objected that (\ref{deltaENRparametrization}) can be ruled out
for macroscopic objects when $\xi_1 \sim 1$. One has $p^{2}/2m\lesssim
\delta E$ for $p\lesssim p_{0}\equiv \xi_1 m^{2}c/M_{P}$ so that the
deformation $\delta E$ dominates over the Minkowski kinetic term.
Standard-model particles with $m\lesssim 10^{-16}M_{P}$ makes $\delta E$
dominate in the extreme NR limit $p\lesssim p_{0}\sim 10^{-16} \xi_1 mc$.
However, for macroscopic objects one can easily have $m\sim M_{P}$ and so
that $p^{2}/2m\lesssim \delta E$ for $p\lesssim p_{0}\sim \xi_1 mc$, i.e.,
the deformation $\delta E$ dominates in the entire NR regime. But this would
contradict the familiar dynamics of classical NR bodies. Note, however, that
(\ref{deltaENRparametrization}) is merely the small-$p$ asymptotic expansion
of the full deformation $\delta E(p,m,M_{P})$, and is thus valid for all
momenta up to some $p_{\lambda }$, where $p_{\lambda }$ depends on the
explicit functional form of $\delta E$. For example a deformation

\begin{equation}
\delta E(p,m,M_{P},p_{\lambda })=\xi _{1}\frac{mc\,p}{2M_{P}}\exp
(-p/p_{\lambda })  \label{example}
\end{equation}%
behaves as $\delta E=\xi _{1}mc\,p/2M_{P}$ for $p\lesssim p_{\lambda }$ and $%
\delta E\simeq 0$ for $p\gg p_{\lambda }$. Therefore, to apply (\ref{example}%
) to macroscopic bodies one should measure the momenta of extended objects
with $p\leq p_{\lambda }$ and this is impossible for sufficiently small $%
p_{\lambda }$ below the lowest measurable momentum for extended bodies.
Since one supposes that this is always the case for $p_{\lambda }$, the
relation (\ref{deltaENRparametrization}) cannot be ruled out for classical
macroscopic bodies. We also emphasize how (\ref{deltaENRparametrization}) is
commonly accepted in the literature \cite{amelino,am1}{\large \textbf{%
. }}

This Letter addresses the effect of a deformed dispersion relation
on the critical temperature $T_{c}$ of a spatially-uniform
non-trapped Bose-Einstein condensate (BEC). This effect can be
compared with the current precision in $T_c$ measurements that
allows constraining the leading order parameter of the NR
deformation (\ref{deltaENRparametrization}) up to $|\xi_1|
\lesssim 10^2$.

After languishing for seven decades as a mere academic exercise in
textbooks, BEC was finally observed in the laboratory in laser-cooled,
magnetically-trapped ultracold bosonic clouds of $_{37}^{87}Rb$ atoms \cite%
{Ander},$\ _{3}^{7}Li$ \cite{Bradley}, $_{11}^{23}Na$ \cite{Davis}, $%
_{1}^{1}H$ \cite{Fried}, $_{37}^{85}Rb$ \cite{Cornish}, $_{2}^{4}He$ \cite%
{Pereira}, $_{19}^{41}K$ \cite{Mondugno}, $_{55}^{133}Cs$ \cite{Grimm}, $%
_{70}^{174}{Yb}$ \cite{Takasu03} and $_{24}^{52}Cr$
\cite{Griesmaier}. The relativistic BEC \textit{including
}antibosons as well as bosons has been reported \cite{PRL07Baker},
but for simplicity we neglect antibosons even in the UR limit in
what follows. A generalization including antibosons is
straightforward.  Nowadays there exist very accurate measurements
of BEC critical temperatures $T_{c}$\ so that the aforementioned
constraints can be addressed. Specifically, in Ref.
\cite{condensatePRL} very accurate measurements of $T_{c}$ shifts
due to strong interboson interactions are reported for
$_{19}^{39}K$. Such high-precision measurements might be useful
in constraining Planck-scale dispersion relation deformations introduced in (%
\ref{deformedenergydensitydefinition0}) in feasible experiments.

Neglecting interboson interactions in the BEC we find that the leading-order
deformation in (\ref{deltaENRparametrization}) produces a shift $\Delta
T_{c}/T_{c}^{0}\propto \xi _{1}m^{2}c/\hbar M_{P}n^{1/3}$ where $\Delta
T_{c}\equiv T_{c}-T_{c}^{0}$ with $\,T_{c}$ the critical temperature of the
gas with the deformed dispersion relation, $T_{c}^{0}$ that same temperature
in the undeformed Minkowski case, $n$ the boson number density and $m$ the
boson mass. This shift can be unexpectedly high. For example, we obtain $%
\Delta T_{c}/T_{c}^{0}\sim 10^{-4}\,\xi _{1}$ for $_{37}^{85}Rb$
with a particle number density $n\simeq 10^{12}\,cm^{-3}$.  Since
in high-precision measurements of $\Delta T_{c}/T_{c}^{0}$ in
$^{39}_{19}K$ due to interboson interactions is $\Delta
T_{c}/T_{c}^{0} \sim 5\times 10^{-2}$ within a $1\%$ error
\cite{condensatePRL}, the deformation parameter can be constrained
up to $|\xi_1|\lesssim 10^2$. Moreover, since $\Delta
T_{c}/T_{c}^{0}\propto n^{-1/3}$ one could enlarge the temperature
shift merely by reducing $n$ (without however making it so small
that it invalidates the thermodynamic limit), and therefore the
bound on the deformation parameter $\xi_1$ can be improved using
dilute (small $n$) BECs.

Based on such an unexpected relevant result, it behooves one to
propose experiments with dilute BECs with the aim of measuring a
Planck-scale induced shift of $T_{c}$. Of course, one should
eventually generalize our result concerning $\Delta
T_{c}/T_{c}^{0}$ to trapped BECs but this is not trivial. The main
task there would be to calculate the energy levels of the trapped
bosons that in the case of free bosons corresponds to the deformed
dispersion relation (\ref{deltaENRparametrization}), with the
result expected to depend on the specific quantum-gravity
framework used. Additionally, in planning a feasible experiment
one should include effects currently measurable experimentally
\cite{condensatePRL} due to interboson interactions and compare
them Planck-scale contributions to $\Delta
T_{c}/T_{c}^{0}$.\textbf{\ }These caveats are currently under
scrutiny \cite{PRDManuel} (see also
refs.\cite{castellanos1,castellanos2}).

We also analyze the effect of the next-to-leading-term deformation in (\ref%
{deltaENRparametrization}) and show that one obtains $\Delta
T_{c}/T_{c}^{0}=\xi _{2}m/M_{P}$ which, being an extremely small and
constant shift, cannot constrain the deformation parameter $\xi _{2}$ with $%
T_{c}$ measurements alone. Therefore, the next-to-leading-order deformation
in (\ref{deltaENRparametrization}) cannot be excluded nor even bounded.
Lastly, we derive the effect of a Planck-scale dispersion relation
deformation in the UR limit (\ref{deltaEURparametrization}) and find the
remarkable result that this effect is always negligibly small. All this
suggests that Planck-scale corrections to $T_{c}$ can be important for NR
BECs rather than UR ones.

The general procedure to calculate $\Delta T_{c}/T_{c}^{0}$ is first
sketched. We write the general form
\begin{equation}
\delta E(p)=xf(p)  \label{deformedcasex}
\end{equation}%
for the dispersion relation deformation $\delta E(p)$ where $x\ll
1$ is a dimensionless deformation parameter. For the
dispersion-relation deformation defined in (\ref{deformedcasex})
the condensation temperature $T_{c}(x)$ is a function of the
deformation parameter $x$ and is obtained by extracting $T_{c}(x)$
from its defining implicit relation
\begin{equation}  \label{Iintegral}
2\pi ^{2}\hbar ^{3}n=\int_{p_0}^{\infty }\left[ \exp \left[ \frac{%
E_{0}(p)+x\,f(p)-mc^{2}}{k_{B}T_{c}(x)}\right] -1\right]
^{-1}p^{2}dp. \label{numericaltc}
\end{equation}
where $p_0 \equiv  \pi \hbar/L$  for free particles in a box of
volume $L^3$. Clearly, for $x=0$ one recovers the usual Minkowski
value $T_{c}(0)=T_{c}^{0} $ that in the NR limit $E_{0}(p)\simeq
mc^{2}+p^{2}/2m$ gives the familiar BEC formula
\begin{equation}  \label{T0}
T_{c}^{0}=T_{c}^{NR}\equiv \frac{2\pi }{\zeta (3/2)^{2/3}}\frac{\hbar
^{2}n^{2/3}}{k_{B}m}.  \label{TMinkowskian}
\end{equation}%
Since the lhs of (\ref{numericaltc}) is independent of $x$ one has $\partial
_{x}n=0$. After some algebra one readily obtains

\begin{equation}
\begin{array}{ll}
\partial _{x}T_{c}(x)/T_{c}(x)=\int_{p_0}^{\infty
}f(p)g(p,x)p^{2}dp/\int_{p_0}^{\infty }p^{2}dp\times
\\
\\
\left[ E_{0}(p)+x\,f(p)-mc^{2}\right] g(p,x)  \label{deltaTsuT1}
\end{array}
\end{equation}
where we have defined

\begin{equation}
\begin{array}{ll}
g(p,x)\equiv \left( \exp \left[ \frac{E_{0}(p)+x\,f(p)-mc^{2}}{k_{B}T_{c}(x)}%
\right] -1\right) ^{-2}\times \\
\\
\exp \left[ \frac{E_{0}(p)+x\,f(p)-mc^{2}}{k_{B}T_{c}(x)}\right] .
\end{array}
\end{equation}
This expression is quite useful to calculate the shift in $T_{c}$ due to the
dispersion relation deformation. In fact, since $x\ll 1$, one can take%
\begin{equation}
\frac{\Delta T_{c}}{T_{c}^{0}}=\frac{T_{c}(x)-T_{c}(0)}{T_{c}(0)}\simeq
x\left( \frac{\partial _{x}T_{c}(x)}{T_{c}(x)}\right) _{|_{x=0}}
\label{temperatureshift}
\end{equation}%
and the last term can be evaluated by use of (\ref{deltaTsuT1}). Let us
consider the leading term of the NR deformation given in (\ref%
{deltaENRparametrization}), i.e., $\delta E=\xi _{1}mcp/2M_{P}$. This
corresponds to $x=\xi _{1}m/2M_{P}$ and $f(p)=cp$. In the NR limit one can
write $E_{0}(p)\simeq mc^{2}+p^{2}/2m$ and $T_{c}(0)=T_{c}^{NR}$, so that (%
\ref{temperatureshift}) becomes

\begin{equation}
\frac{\Delta T_{c}}{T_{c}^{0}}\simeq  0.1 \, \left(
\frac{m^{2}c}{\hbar M_{P}\,n^{1/3}}\right)  \, \xi _{1} \, \ln(N).
\label{deltaTsuT3}
\end{equation}
where $N \equiv n L^3$ is the total number of particles. This
shift can be evaluated for a $_{37}^{85}Rb$ BEC\ with number
density $n\simeq 10^{12}cm^{-3}$ \cite{dellano2}, boson mass
$m\simeq 150\times 10^{-27}kg$ and $N=10^9$. Hence $\Delta
T_{c}/T_{c}^{0}\simeq 8.6\times 10^{-5}\xi _{1}$. Since one
expects $\xi _{1}\sim 1$ in quantum-gravity theories, this can be
extremely large when compared to the strength of the deformation in (\ref%
{deltaENRparametrization}) which is of order $\delta E/E \simeq \xi_1 p/2 c
M_{P} \ll \xi_1 m/M_p \sim 10^{-17}$. Evidently, the temperature shift (\ref%
{deltaTsuT3}) being $\varpropto $ $n^{-1/3}$ can be enlarged for
BECs with sufficiently small $n$, but not so small to render the
thermodynamic limit inapplicable. This fact is vital since one can
then seek low-density BECs with a correspondingly large $\Delta
T_{c}/T_{c}^{0}$\ in order to constrain and perhaps even measure
the deformation parameter $\xi _{1}$. Moreover, from
(\ref{deltaTsuT3}) another way of enlarging $\Delta
T_{c}/T_{c}^{0}$\ is to consider more massive bosons.

As mentioned initially, if one is to deal with actual laboratory
measurements of $\Delta T_{c}/T_{c}^{0}$ one should generalize (\ref%
{deltaTsuT3}) to the case of a trapped gas. However, such a
generalization is a delicate matter well beyond the intent of this
paper and that will be presented elsewhere \cite{PRDManuel}. Here
we merely stress that, for non-trapped BECs the shift $\Delta
T_{c}/T_{c}^{0}$\ due to Planck-scale effects can be as large as
$\sim 10^{-4}\xi _{1}$ and even larger for dilute BECs. Since we
expect a comparable effect for trapped
BECs, it makes sense to compare such a shift with empirical values of $%
\Delta T_{c}/T_{c}^{0}$.

In high-precision measurements of $\Delta T_{c}/T_{c}^{0}$\ in
$_{19}^{39}K$ \cite{condensatePRL} due to interboson interactions,
the order of magnitude is $\Delta T_{c}/T_{c}^{0}\simeq 5\times
10^{-2}$ with at most a $1\%$ error so that such measurements may
be sensitive to Planck-scale effects. As seen above, the
Planck-scale-induced shift in the condensation temperature is
$\Delta T_{c}/T_{c}^{0}\simeq 10^{-4}\,\xi _{1}$ for a $
_{37}^{85}Rb$ BEC, which allows to constrain the deformation
parameter up to $|\xi_1|\lesssim 10^2$. Moreover, since the
temperature shift can be enhanced for even more dilute and/or more
massive BECs, even better bounds on $\xi_1$ are obtainable.

Note also that (\ref{deltaTsuT3}) may theoretically suggest the exclusion of
the leading term in (\ref{deltaENRparametrization}) as it would cause an
unbounded shift for very small $n$. Even though the temperature shift
becomes arbitrarily large for small $n$ the critical temperature vanishes as
$n^{1/3}$ when $n\rightarrow 0$, since one has

\begin{equation}
T_{c}\simeq \frac{2\pi }{\zeta (3/2)^{2/3}}\frac{\hbar ^{2}n^{2/3}}{k_{B}m}%
\left[ 1+30\,\left( \frac{m^{2}c}{\hbar M_{P}\,n^{1/3}}\right) \,\xi _{1}%
\right] .
\end{equation}%
Thus, a nonzero deformation parameter $\xi _{1}$ is not inconsistent.
Remarkably therefore, a measured nonzero shift of $T_{c}$ in a low-density
BEC, but unrelated to interboson interactions, would imply a leading-order
dispersion relation deformation.

Let us briefly examine the next-to-leading-order term in
(\ref{deltaENRparametrization}), corresponding to $\xi _{2}\neq 0$
and $\xi _{1}=\xi _{3}=0$ with the bound $|\xi _{2}|\leq 10^{9}$
\cite{amelino,am1}. This implies the energy dispersion relation
\begin{equation}
E\simeq mc^{2}+p^{2}/2m+\xi _{2}p^{2}/2M_{P}\equiv
mc^{2}+p^{2}/2m^{\xi _{2}}
\end{equation}
where $m^{\xi _{2}}\equiv mM_{P}/(\xi _{2}m+M_{P})$. This gives $
T_{c}=T_{c}^{0}\left[ 1+\xi _{2}(m/M_{P})\right] $ so that the
temperature shift becomes
\begin{equation}
\frac{\Delta T_{c}}{T_{c}^{0}}=\xi _{2}\frac{m}{M_{P}}.
\end{equation}
Since $m/M_{P}\ll 1$\ such a shift is extremely small and therefore the
next-to-leading term in (\ref{deltaENRparametrization}) cannot be excluded
nor bounded even for very large values of $\xi _{2}$.

So far our focus has been limited to the NR limit of the deformed
dispersion relation (\ref{deltaENRparametrization}). We now show
that Planck-scale corrections to $T_{c}$ are negligibly small as a
result of the leading term in the UR limit
(\ref{deltaEURparametrization}). In this case $\delta E(p)=\eta
_{1}p^{2}/2M_{P}$ and the temperature shift can be calculated via
( \ref{deltaTsuT1}-\ref{temperatureshift}) with $x=\eta
_{1}m/2M_{P}$ and $f(p)=p^{2}/m$. Moreover, one has
$E_{0}(p)\simeq cp$ and $ T_{c}(0)=T_{c}^{UR} $ where $T_{c}^{UR}$
is the UR condensation temperature
\begin{equation}
T_{c}^{UR}\equiv \hbar \,c\,\pi ^{2/3}n^{1/3}/k_{B}\zeta (3)^{1/3}.
\label{TUR}
\end{equation}
The resulting temperature shift is thus
\begin{equation}
\frac{\Delta T_{c}}{T_{c}^{0}}\simeq 1.8\,\eta _{1}\left( \frac{
k_{B}T_{c}^{UR}}{c^{2}M_{P}}\right)  \label{temperatureshiftUR}
\end{equation}
and such a shift can be appreciable only for extremely high densities $n$
such that $k_{B}T_{c}^{UR}\sim c^{2}M_{P}\sim 10^{19}GeV$. This result is
significant as being counterintuitive since one expects that Planck-scale
effects be appreciable in UR rather than NR phenomena. In fact, the
situation is quite the opposite: Planck-scale corrections to BEC critical $%
T_{c}$s are appreciable in NR BECs (\ref{deltaTsuT3}) but are extremely
small for UR BECs (\ref{temperatureshiftUR}).

Lastly, we briefly discuss the cosmological consequences of our
analysis. It was recently proposed
\cite{cosmologicalcondensate,c1,c2,c3,c4,c5,c6,c7,c8,c9,c92,c93,c10,c11,c12,c13,c14,c15,c16,c17,c18,c19,briscese}
that dark matter in the universe might consist of a BEC phase due
to some boson. Since the condensation temperature determines the
epoch of formation of such a condensate, a shift in $T_c$ may
delay or anticipate the condensation of dark-matter particles,
thus affecting the phenomenology of the model. This motivates
exploring whether Planck-scale deformation can in fact affect
BECs. At any rate, one is far from a direct observation of such a
cosmological condensate so that at present its existence is
speculative, even though one can search for its indirect
cosmological traces. Indeed, this suggests an interesting avenue
of research.

To conclude, we have determined the effects of Planck-scale
deformation of the dispersion relation on the condensation
temperature of a non-trapped BEC. In particular, assuming a
nonrelativistic leading-order dispersion relation defined in
(\ref{deltaENRparametrization}) one finds that such an effect may
be comparable with the precision measurement of $T_{c}$
\cite{condensatePRL} and that one can bound the leading-order
deformation parameter up to $|\xi_1|\lesssim 10^2$. We thus argue
that one should generalize this result to the case of trapped BECs
\cite{PRDManuel} in order to propose feasible experiments
sensitive to Planck-scale physics.\textbf{\ }We
have also shown that the next-to-leading correction in (\ref%
{deltaENRparametrization}) causes an extremely small shift that is
unobservable with any experiment so such a deformation can neither be
excluded nor fixed outside presently existing bounds. Finally, it is
noteworthy that Planck-scale corrections to $T_{c}$ are extremely small for
UR condensates.
This leads one to the conclusion that Planck-scale physics may be relevant
for cold-atom NR BECs rather than UR BECs.

\textbf{Acknowledgements}: FB thanks G. Amelino-Camelia for useful
discussions during the early stages of this Letter and for his continued
encouragement since then. It was completed during a visit of FB at IIM-UNAM
in Mexico City. FB and MdeLl thank UNAM-DGAPA-PAPIIT (M\'{e}xico) for its
support from grant IN106908.

\end{document}